\documentclass[twocolumn,prl,superscriptaddress]{revtex4}
\usepackage{amsmath,amssymb,amsfonts,mathrsfs}
\usepackage{amsthm}
\usepackage{graphicx}
\usepackage{dcolumn}
\usepackage{bm}
\usepackage{color}

\begin{document}

\title{Topological Maxwell Metal Bands in a Superconducting Qutrit}

\author{Xinsheng Tan}
\affiliation{National Laboratory of Solid State Microstructures, School of Physics,
Nanjing University, Nanjing 210093, China}
\author{Dan-Wei Zhang}
\affiliation{Guangdong Provincial Key Laboratory of Quantum Engineering and Quantum
Materials, SPTE, South China Normal University, Guangzhou 510006, China}
\author{Qiang Liu}
\affiliation{National Laboratory of Solid State Microstructures, School of Physics,
Nanjing University, Nanjing 210093, China}
\author{Guangming Xue}
\affiliation{National Laboratory of Solid State Microstructures, School of Physics,
Nanjing University, Nanjing 210093, China}
\author{Hai-Feng Yu}
\email{hfyu@nju.edu.cn}
\affiliation{National Laboratory of Solid State Microstructures, School of Physics,
Nanjing University, Nanjing 210093, China}
\author{Yan-Qing Zhu}
\affiliation{National Laboratory of Solid State Microstructures, School of Physics,
Nanjing University, Nanjing 210093, China}
\author{Hui Yan}
\affiliation{Guangdong Provincial Key Laboratory of Quantum Engineering and Quantum
Materials, SPTE, South China Normal University, Guangzhou 510006, China}
\author{Shi-Liang Zhu}
\email{slzhu@nju.edu.cn}
\affiliation{National Laboratory of Solid State Microstructures, School of Physics,
Nanjing University, Nanjing 210093, China}
\affiliation{Guangdong Provincial Key Laboratory of Quantum Engineering and Quantum
Materials, SPTE, South China Normal University, Guangzhou 510006, China}
\affiliation{Synergetic Innovation Center of Quantum Information and Quantum Physics,
University of Science and Technology of China, Hefei, Anhui 230026, China}
\author{Yang Yu}
\email{yuyang@nju.edu.cn}
\affiliation{National Laboratory of Solid State Microstructures, School of Physics,
Nanjing University, Nanjing 210093, China}
\affiliation{Synergetic Innovation Center of Quantum Information and Quantum Physics,
University of Science and Technology of China, Hefei, Anhui 230026, China}

\begin{abstract}

We experimentally explore the topological Maxwell metal bands by mapping the momentum space of condensed-matter models to the tunable parameter space of superconducting quantum circuits. An exotic band structure that is effectively described by the spin-1 Maxwell equations is imaged. Three-fold degenerate points dubbed Maxwell points are observed in the Maxwell metal bands. Moreover, we engineer and observe the topological phase transition from the topological Maxwell metal to a trivial insulator, and report the first experiment to measure the Chern numbers that are higher than one.
\end{abstract}

\maketitle

\bigskip

Discovery of new states of matter in condensed-matter materials or synthetic
systems is at the heart of modern physics \cite{Castro,Hasan,Qi}. The last
decade has witnessed a growing interest in engineering quantum systems with
novel band structures for topological states, ranging from graphene and
topological insulators \cite{Castro,Hasan,Qi,Tarruell,Zhu,Bloch2015,Duca},
to topological semimetals and metals \cite{Xu1,Lv1,Wan,Lu,Chan}. The band
structures of graphene and some topological insulators or semimetals near
the two-fold degenerate points simulate relativistic spin-1/2 particles in
the quantum field theory described by the Dirac or Weyl equation.
Most interestingly, the Dirac and Weyl bands have rich topological
features \cite{Castro,Hasan,Qi,Tarruell,Zhu,Duca,Bloch2015,Xu1,Lv1,Wan,Lu,Chan}.
For instance, states in the vicinity of a Weyl point possess a
non-zero topological invariant (the Chern number). The topological
bands with two-fold degenerate points so far realized simulate
conventional Dirac-Weyl fermions studied in the quantum field
theory. On the other hand, unconventional bands with topological properties
mimicking higher spinal counterparts are also fundamentally
important but rarely studied in condensed matter physics or
artificial systems \cite{Bradlyn,Liang}, noting that they provide potentially a quantum family to find
quasiparticles that have no high-energy analogs, such as integer-(speudo)spin fermionic excitations. Recently, a piece of pioneering work
in this direction theoretically predicted that new fermions beyond
Dirac-Weyl fermions can emerge in some band structures with three-
or more-fold degenerate points \cite{Bradlyn}. The three-fold degeneracies in the bands carry
large Chern numbers $C=\pm2$ and give rise to two chiral Fermi arcs and the spin-1 quasiparticles \cite{Bradlyn}. The spin-1 particles can exhibit striking relativistic quantum dynamics beyond the Dirac dynamics \cite{Castro}, such as super-Klein tunneling and supercollimation effects \cite{Fang} and geometrodynamics of spin-1 photons \cite{Bliokh}. However, these topological bands with the unconventional fermions have yet been observed in real materials or artificial systems. Several challenges may hinder their experimental
investigation in conventional materials and condensed matter systems. The first is that the realization of
the spin-1 Hamiltonian requires unconventional spin-orbital interactions in
three-dimensional (3D) periodic lattices \cite{Bradlyn}. Second, it is difficult to continuously tune the parameters in
materials to study fruitful topological properties including topological transition. Moreover, it is difficult to
directly detect the topological invariant of the multi-fold degenerate points in condensed
matter systems. Nevertheless, artificial superconducting quantum circuits possessing high controllability \cite{You,Abdumalikov,Tan,Ray,Nakamura,Siddiqi,Wal,Feofanov,Corcoles,Deng,Kim,Vion,Sun,Schroer,Roushan}
provide an ideal and powerful tool for quantum simulation and
the study of novel quantum systems, including the topological ones
\cite{Schroer,Roushan}.

In this Letter, we experimentally explore
an unconventional topological band structure, called Maxwell
metal bands, with a superconducting qutrit via an analogy between
the momentum space of the presented condensed-matter model and the
tunable parameter space of superconducting quantum circuits. By
measuring the whole energy spectrum of our system, we clearly
image a new band structure, which consists of a flat band and two three-fold degenerate points in the 3D
parameter space dubbed Maxwell points. The system dynamics
near the Maxwell points are effectively described by
the analogous spin-1 Maxwell equations. We further investigate the
topological properties of Maxwell metal bands by measuring the
Chern numbers $\pm2$ of the simulated Maxwell points from the
non-adiabatic response of the system.
By tuning the Hamiltonian parameters, we engineer the topological phase
transition from the Maxwell metal to a trivial insulator, which is
demonstrated unambiguously from the evolution of tunable band structures and
Chern numbers across the critical points.

\begin{figure}[tbp]
\centering
\includegraphics[width=8cm]{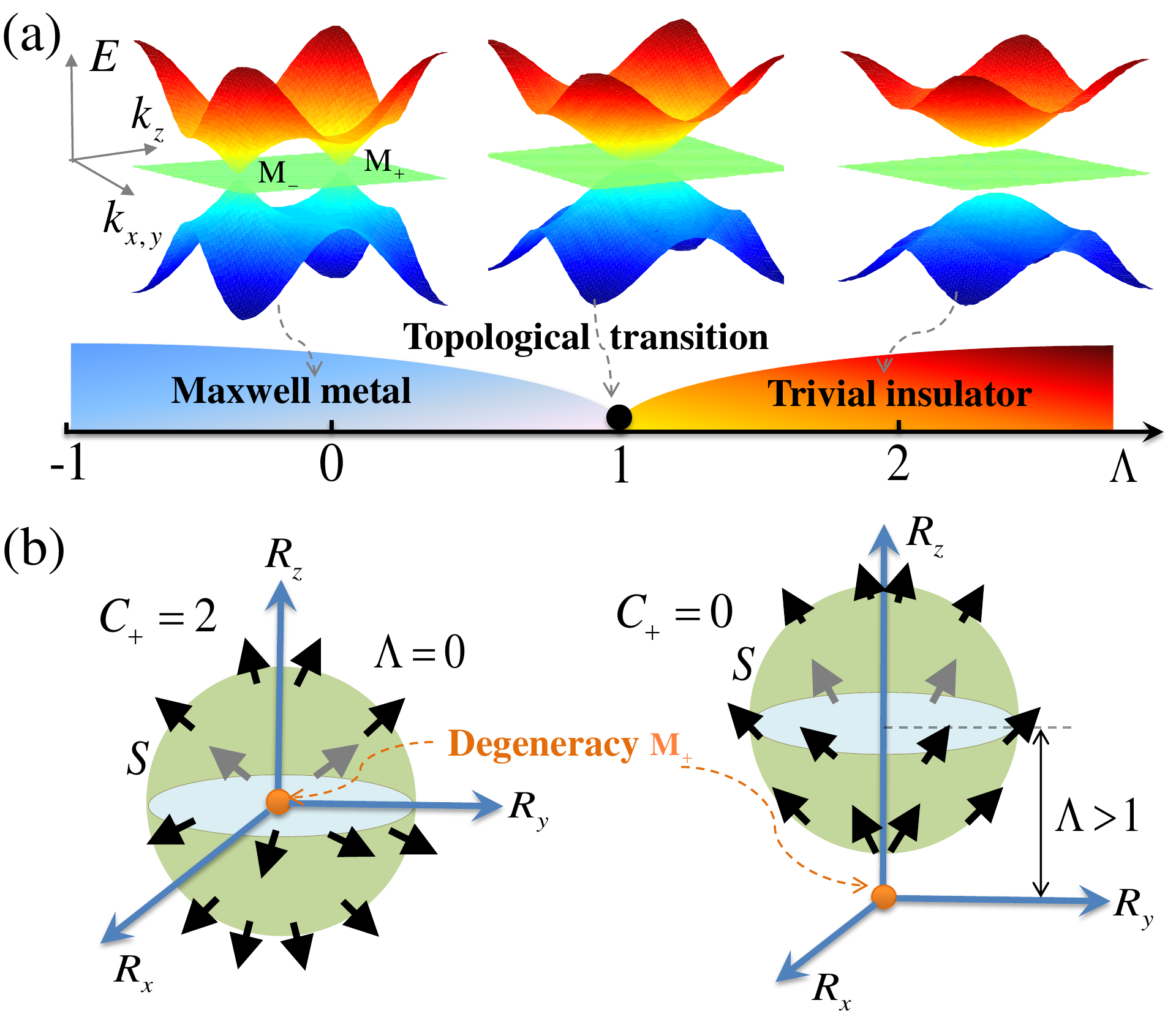}
\caption{(Color online) Phase diagram and geometric illustration of
the spin-1 Maxwell system. (a) Phase diagram of the Maxwell system
with respect to the parameter $\Lambda$. From left to right: the energy spectra for the Maxwell
metal phase with a pair of Maxwell points denoted by $\mathbf{M}_{\pm}$ ($\Lambda=0$), the topological transition point with the merging of the two
points ($\Lambda=1$), and the trivial insulator phase with band gaps ($\Lambda=2$). (b) Geometric illustrations
of the topological difference between the two distinct phases when the spherical manifold $\mathcal{S}$ moves from the degeneracy in the
$z$ direction by distance $\Lambda$. The Berry flux vectors are
schematically presented by arrows, showing the different signature
textures for the topological and trivial phases: the vectors fully (do not) wind
around in the topological (trivial) case with $\Lambda=0$ ($\Lambda>1$), giving the Chern number $C_+=2$ ($C_+=0
$).}
\label{PhaseDiagram}
\end{figure}

We realize the following model Hamiltonian in momentum space describing a free pseudospin-1 particle \cite{Supp}
\begin{equation}  \label{BlochHam}
\mathcal{H}(\mathbf{k})=R_xS_x+R_yS_y+R_zS_z,
\end{equation}
where $\mathbf{k}=(k_x,k_y,k_z)$ denote the quasimomenta, $R_x=\sin{k_x},
R_y=\sin{k_y}$, and $R_z=\Lambda+2-\cos{k_x}-\cos{k_y}-\cos{k_z}$ are the
Bloch vectors with a control parameter $\Lambda$, and $S_{x,y,z}$ are the spin-1 matrices. The resulting three bands can touch at certain points in
the first Brillouin zone for proper $\Lambda$ with a zero-energy flat band
in the middle. For instance, when $|\Lambda|<1$, the bands host two
threefold degeneracy points at $\mathbf{M}_{\pm}=%
\begin{pmatrix}
0,0,\pm\arccos\Lambda%
\end{pmatrix}%
$. 
Near $\mathbf{M}_{\pm}$ one has the low-energy effective Hamiltonian
\begin{equation}  \label{MWHam}
\mathcal{H}_{{\pm}}(\mathbf{q})=q_{x}S_{x}+q_{y}S_{y}\pm{\alpha q_{z}S_{z}},
\end{equation}
with $\alpha=\sqrt{1-\Lambda^2}$ and
$\mathbf{q}=\mathbf{k}-\mathbf{M}_{\pm}$ for the two degeneracy
points. Equation (2) is analogous to the Maxwell Hamiltonian for photons
and the dynamics of the low-energy pseudospin-1 excitations are
effectively described by the Maxwell equations \cite{YQZhu,Stone}. In
this sense, the threefold degeneracy points are named Maxwell
points, similar to the Dirac and Weyl points in some
(pseudo)spin-1/2 systems, such as graphene and Weyl semimetals
\cite{Castro,Hasan,Qi,Tarruell,Zhu,Duca,Bloch2015,Xu1,Lv1,Wan,Lu}.

The spin-1 system described by Hamiltonian (\ref{BlochHam}) has two
different topological phases determined by the parameter $\Lambda$: the
topological Maxwell metal phase with a pair of Maxwell points in the bands
when $|\Lambda|<1$ and the trivial insulator phase with band gaps when $%
|\Lambda|>1$. At the critical point $\Lambda=1$, the two Maxwell points
merge and then disappear at the band center, indicating the topological
transition. The phase diagram and typical band structures are illustrated in
Fig. \ref{PhaseDiagram}(a) ($\Lambda=0,1,2$ from left to right). The
topological nature of Maxwell metal bands can be revealed from the two
Maxwell points acting as the sink and source of the Berry flux in 3D momentum or
parameter space. Moreover, the topological invariant of the Maxwell points $\mathbf{M}_{\pm}$ is given by the Chern numbers $C_{\pm}$ defined as the integral over a closed manifold $\mathcal{S}$ (contains the equivalent energy points of $\mathcal{H}_{{\pm}}$) enclosing each of the points in the momentum or parameter space of $\mathcal{H}_{{\pm}}$
\begin{equation}
C_{\pm}=\frac{1}{2\pi}\oint_{\mathcal{S}}\mathbf{F}_{\pm}\cdot d\mathbf{S}=\pm2,
\end{equation}
where $\mathbf{F}_{\pm}$ denote the vector form of the Berry curvature
\cite{Supp}. Hence, the transition between the two distinct phases
can be topologically represented by the movement of a spherical manifold $\mathcal{S}$ of radius $1$ from the degeneracy in the $z$ direction by distance $\Lambda$, as shown in Fig. \ref{PhaseDiagram}(b). When $|\Lambda|<1$ the degeneracy
lies within $\mathcal{S}$, giving $C_{\pm}=2$ for the Maxwell metal phase; when $|\Lambda|>1$ it lies outside $\mathcal{S}$, giving $C_{\pm}=0$ for the
trivial insulator phase.


\begin{figure}[tbph]
\includegraphics[width=8cm]{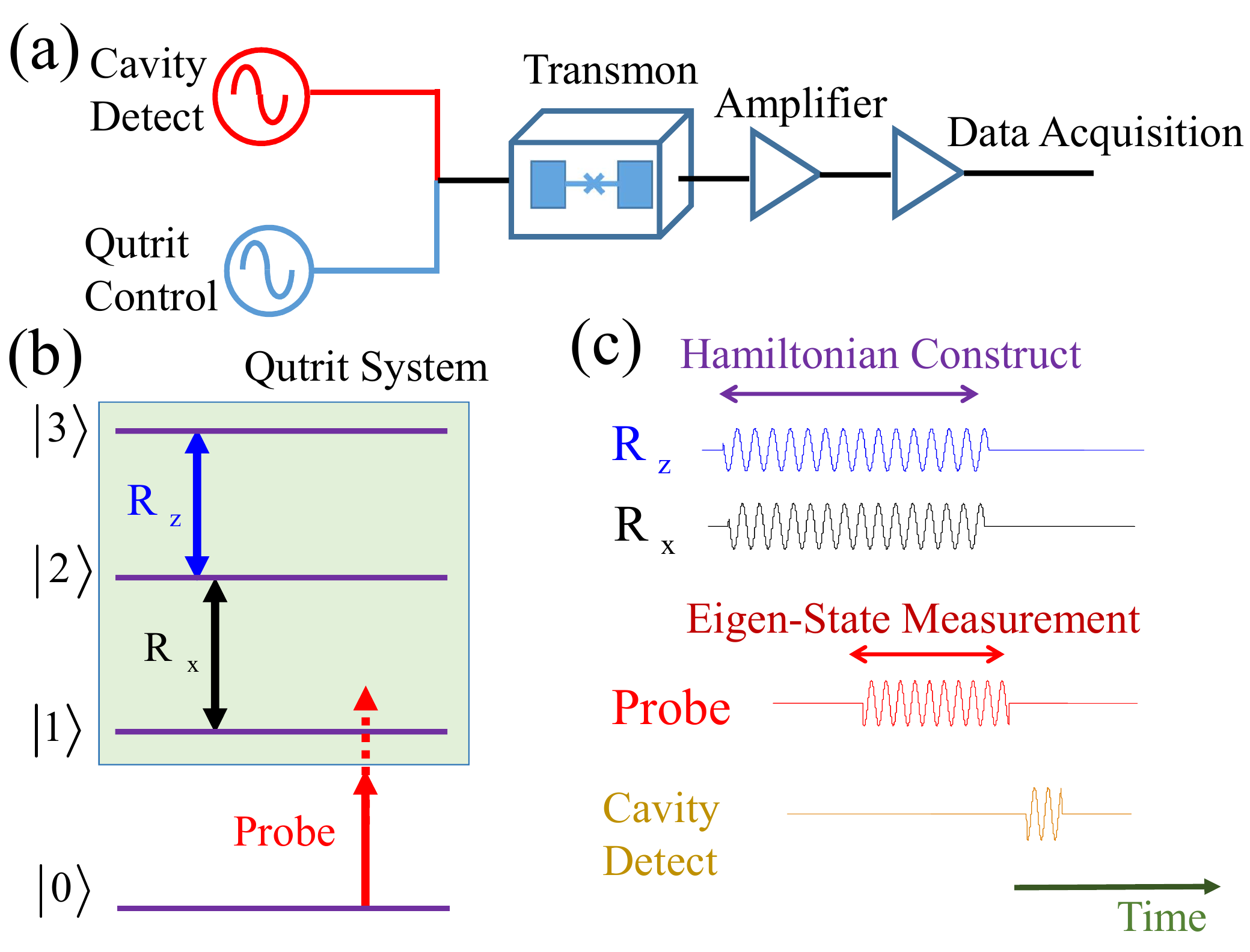}
\caption{(Color online) (a) Schematic of experimental
setup for controlling and measuring a 3D superconducting transmon qutrit. The microwaves for manipulating and measuring the qutrit are
applied to the sample. We use digital heterodyne for data acquisition.
Amplifiers and filters are used to increase signal-to-noise ratio and
isolate qutrit from external noise. (b) Schematic of the relevant energy levels of the
transmon for spectroscopy measurement. Levels $|1\rangle,
|2\rangle$, and $|3\rangle$ form the spin-1 basis $\{(1,0,0)^{\text{T}}, (0,1,0)^{\text{T}},(0,0,1)^{\text{T}}\}$. $
|0\rangle $ serves as the reference level to measure the spectroscopy by sweeping the frequency of the probe microwave (schematically illustrated by the dashed arrow). (c) Time profile for spectroscopy measurement. The control microwave pulses $R_x$ and $R_z$ drive the system to form the eigenstates, which are empty. Then the probe microwave pulse pumps the system from $|0\rangle $ to the eigenstates when its frequency matches the level spacing. By sending a detect pulse to the cavity, we can readout the population in the eigenstates as resonant peaks. By collecting the resonant peaks we obtain the energy spectrum of the Maxwell metal.}
\label{Setup}
\end{figure}

Below we simulate the Hamiltonian (\ref{BlochHam}) with a fully controllable artificial
superconducting qutrit. The sample used in our experiment is a 3D transmon, which consists
of a superconducting qutrit embedded in a 3D aluminium cavity
\cite{Paik} of which TE101 mode is at 9.053 GHz. The intrinsic quality factor of the cavity is about $10^6$. The whole sample
package is cooled in a dilution refrigerator to a base temperature
of 30 mK. Fig. \ref{Setup}(a) is a brief schematic of our
experimental setup for manipulating and measuring the 3D transmon
(see Supplementary Information). The principals of manipulation and measurement for a 3D transmon are based on the theory of circuit QED \cite{Blais,Reed}, which describes the interaction of artificial atoms subject to microwave fields. We designed the energy levels of
the transmon to make the system work in the dispersive regime.

\begin{figure}[tbph]
\includegraphics[width=8.5cm]{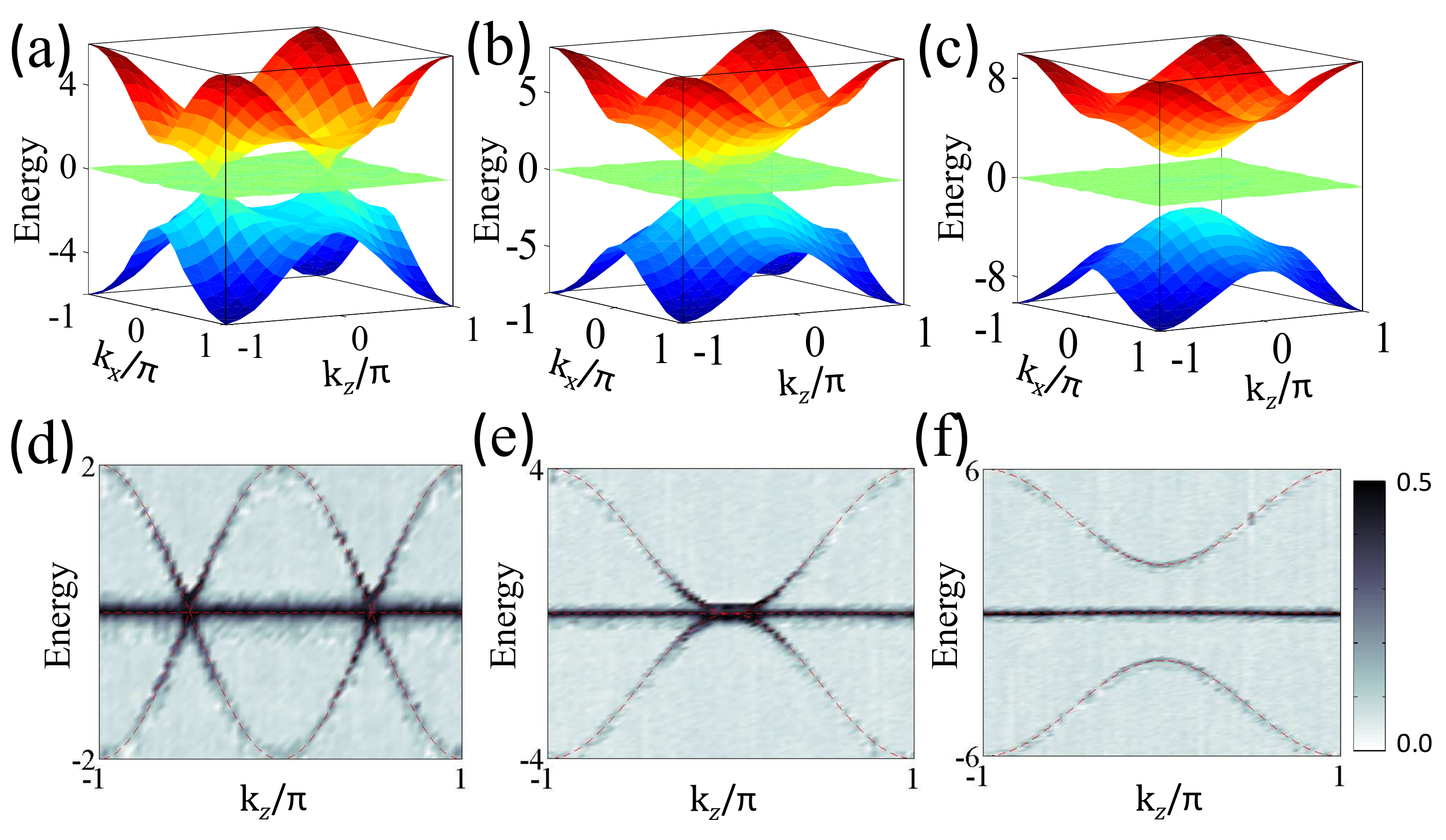}
\caption{(Color online) Measured Maxwell bands. (a), (b), and (c) are band
structures in the first Brillouin zone for $\Lambda=0,1,2$, respectively. (d), (e), and (f) show the corresponding cross sections of band
structures containing Maxwell points in $E-k_{z}$ ($k_x\approx0$) plane of (a) to (c). A linear
dispersion is observed in the Maxwell metal phase. The theoretical calculations are plotted with the red dashed
lines. The spectra are plotted by dropping the overall energy constant $\omega_{01}$ with the energy unit $\Omega=10$ MHz.}
\label{Spectrum}
\end{figure}

In general, the transmon has multiple energy levels and we use the four
lowest ones denoted as $|0\rangle$, $|1\rangle$, $|2\rangle$, and $|3\rangle$. Here $\{|1\rangle, |2\rangle, |3\rangle\}$ form the qutrit
basis and are used to simulate the model Hamiltonian of the Maxwell system,
and $|0\rangle $ is set as a reference level for measuring spectrum [Fig.
\ref{Setup}(b)]. First, we calibrated the transmon. The transition
frequencies between different energy levels are $\omega _{01}/2\pi =$
7.17133 GHz, $\omega _{12}/2\pi =$ 6.8310 GHz and $\omega _{23}/2\pi =$
6.4470 GHz, which are independently determined by saturation spectroscopies.
The energy relaxation times of the transmon are $T_{1}^{01}\sim $ 15 $\mu s$, $T_{1}^{12}\sim $ 12 $\mu s$ and $T_{1}^{23}\sim $ 10 $\mu s.$ The
dephasing times are $T_{2}^{\ast 01}\sim $ 4.3 $\mu s$, $T_{2}^{\ast 12}\sim
$ 3.5 $\mu s$ and $T_{2}^{\ast 23}\sim $ 3.0 $\mu s$. In order to obtain the
spin-1 Hamiltonian [Eq. (\ref{Ham})], as shown in Fig. \ref{Setup}(b), we apply
microwave fields with frequencies $\omega _{12}$, $\omega _{13}$ and $\omega
_{23}$ to generate transitions between the three levels, respectively
denoted as $R_{x}$, $R_{y}$ and $R_{z}$. These transitions are equivalent to
the rotations with respect to different axes. IQ mixers combined with
arbitrary wave generators are used to control the amplitude,
frequency, and phase of microwave pulses. For the microwave-driven qutrit
system, the Hamiltonian with a tunable parameter $\mathbf{k}$ under rotating wave
approximation can be written as
\begin{equation}
H(\mathbf{k})=%
\begin{pmatrix}
0 & -i\Omega _{12}/2 & i\Omega _{13}/2 \\
i\Omega _{12}/2 & 0 & -i\Omega _{23}/2 \\
-i\Omega _{13}/2 & i\Omega _{23}/2 & 0%
\end{pmatrix}+\omega_{01}\mathbf{I}_{3\times3}.
 \label{Ham}
\end{equation}%
Here we design the transition rates as $\{\Omega _{12},\Omega _{13},\Omega
_{23}\}=\{R_x,R_y,R_z\}$ to mimic the model Hamiltonian in the parameter space, and $\mathbf{I}_{3\times3}$ is the 3 by 3 unit matrix. 
Diagonalizing $H(\mathbf{k})$ yields three eigenstates $|0_d\rangle $ and $|\pm\rangle$,
with the corresponding eigen energies $E_0=\omega_{01}$ as the flat band and $E_{\pm}=\omega_{01}\pm\sqrt{R_x^2+R_y^2+R_z^2}$ as the upper and lowest bands shown in Fig. \ref{PhaseDiagram}(a).

We directly measure the spectroscopy of the driven transmon and obtain the band
structure of $H(\mathbf{k})$. Without loss of generality, we always set $k_{y}=0$
in the band-structure measurement. For given $k_{x}$ and $k_{z}$, the dressed states under the microwaves are eigenstates $|0_d\rangle$ and $|\pm\rangle$. A probe microwave pulse is used to pump the system from $|0\rangle $ to the
eigenstates and the resonant peaks of microwave absorption are detected \cite{Supp}. By mapping the frequency of the resonant peak as a function of $k_{x}$ and $k_{z}$, we extract the entire
band structure over the first Brillouin zone, as illustrated from Fig. \ref
{Spectrum}(a) to \ref{Spectrum}(c), which agree well with the theoretical
results shown in Fig. \ref{PhaseDiagram}(a). The topological properties of the
spin-1 Maxwell system depend on $\Lambda$. For $\Lambda$ = 0
[Fig. \ref{Spectrum}(a)], the system is in the Maxwell metal phase and two
Maxwell points located at $(0,\pm \pi /2)$ in $k_{x}$-$k_{z}$ plane are
observed. When $\Lambda$ increases to 1 [Fig. \ref{Spectrum}(b)], two Maxwell
points merge at $(0,0)$, indicating the topological phase transition. They then completely
disappear with further increase of $\Lambda$ and the system becomes a trivial insulator, as shown in Fig. \ref
{Spectrum}(c) ($\Lambda = 2$). This phase transition can be observed more clearly from the cross section
of the Maxwell points in the $E$-$k_{z}$ plane with $k_{x}\approx0$, as shown in Figs. \ref
{Spectrum}(d), (e), and (f) (see Supplementary Materials for the discussion about the spectral brightness distribution). The resonant peaks of 1D spectroscopy data directly image the eigenenergy $E_0$ and $E_\pm$. As predicted by the theory, the dispersion evolves from the linear one (where the quasiparticles are relativistic) near the Maxwell points with a flat band to the quadratic one when crossing the transition point $\Lambda=1$.

\begin{figure}[tbph]
\includegraphics[width=8.5cm]{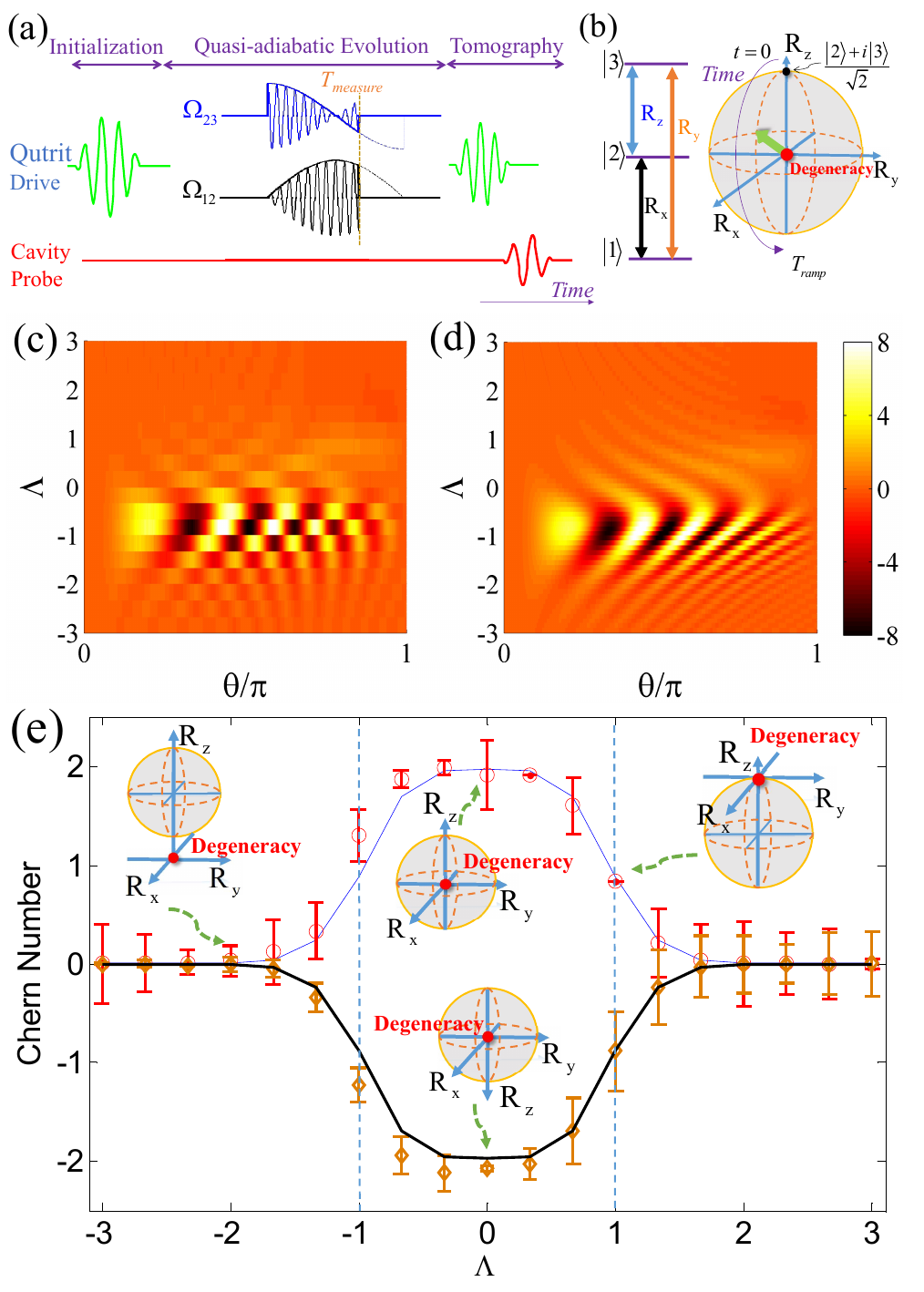}
\caption{(Color online) (a) Time profile for the measurement of
Chern number. The qutrit state is initialized at ($|2\rangle +i|3\rangle $)/$\protect\sqrt{2}$ and then evolves quasi-adiabatically during a non-adiabatic ramp, which is followed by state tomography. (b) Three lowest energy levels $\{|1\rangle, |2\rangle, |3\rangle\}$ coupled by pluses $R_{x,y,z}$ are used to construct the spin-1 Hamiltonian, and the pulse sequence results in a parameter-space motion along the $\phi=0$ meridian ($R_y=0$) on the spherical manifold. (c) and (d) The measured and simulated (with the measured decoherence time of the transmon) Berry curvature $F_{%
\protect\theta \protect\phi}$ as functions of $\protect\theta$ and $\Lambda$. The oscillation pattern suggests a non-adiabatic response. 
(e) The measured (circles and diamonds) and
simulated (solid line) Chern numbers as a function of $\Lambda$ for the Maxwell
points. For $|\Lambda|<1$, $|C_{\pm}|=2$ indicates the Maxwell points in the topological
Maxwell metal phase; for $|\Lambda|>1$, $|C_{\pm}|=0$ indicates the system in the trivial insulator phase.}
\label{Chern}
\end{figure}

We detect the Chern numbers of the Maxwell points by dynamically
measuring the Berry curvature in the parameter space from the non-adiabatic
response in the quasi-adiabatic procedure [Fig. \ref{Chern}(a)]. This non-adiabatic approach \cite{Gritsev} has been shown to be a convenient way to measure the Berry curvature of a spin-1/2 system \cite{Schroer,Roushan}. We here demonstrate that it can be generalized to a spin-1 system. Since the probe level is no longer needed in this measurement, we select
the lowest three levels in the superconducting transmon [Fig. \ref{Chern}(b)],
labeled still as $\{|1\rangle, |2\rangle, |3\rangle\}$ for consistent
definitions in equations. We choose $\{\Omega _{12},\Omega _{13},\Omega _{23}\}=\{\sin\theta \cos\phi,\sin\theta \sin\phi, \cos\theta +\Lambda\}$ to realize the spherical manifold enclosing $\mathbf{M}_{+}$ [Fig. \ref{Chern}(b)], where $\theta\in[0,\pi]$ and $\phi\in [0,2\pi]$ are spherical coordinates. We consider the parameter trajectory that starts at the north pole by preparing the initial qutrit state $(|2\rangle+i|3\rangle )/\sqrt{2}$, which is the eigenstate of the $S_z$ operator. We then linearly ramp the angle $\theta$ as a function $\theta(t)=\pi t/T_{\text{ramp}}$ along the $\phi=0$ meridian in Fig. \ref{Chern}(b). Finally we stop the ramp at various times
$t_{\text{measure}}\in \lbrack0,T_{\text{ramp}}]$ and perform tomography of the qutrit state. In the adiabatic limit, the system
will remain in the meridian. However, we use quasi-adiabatic ramps with fixed $T_{\text{ramp}}=600$ ns $\sim\Omega/10$ ($\Omega=15$ MHz is the energy unit), and the local Berry curvature introduces a deviation from the meridian, which can be defined as the generalized force \cite{Supp} $\langle M_{\phi }\rangle =-\langle \partial _{\phi}H(\theta ,\phi )\rangle|_{\phi=0} =\langle S_{y}\rangle\sin\theta$. Then at each $t_{\text{measure}}$, we
extract the Berry curvature $F_{\theta \phi}\approx\langle M_{\phi}\rangle/v_{\theta}$ from the measured values of $\langle S_{y}\rangle$, where $v_{\theta}=\pi /T_{\text{ramp}}$ is the ramp velocity. As the Hamiltonian is cylindrically invariant around the $z$ axis, a line integral is sufficient for measuring the surface integral of the Chern number as $C=\int_{0}^{\pi }F_{\theta \phi}d\theta$. We extract $F_{\theta\phi}$ of the two Maxwell points $\mathbf{M}_{\pm}$ for $\Lambda=0$ and obtain the Chern numbers $C_+=1.98\pm0.34$ and $C_-=-2.14\pm0.05$, which are close to the theoretical values $\pm2$.

To investigate the topological phase transition in the transmon, we measure $%
F_{\theta\phi}$ of $\mathbf{M}_{\pm}$ as a function of $\theta$ and the tunable parameter $\Lambda$. The measured $F_{\theta\phi}$ of $\mathbf{M}_{+}$ [Fig. \ref{Chern}(c)] is in good agreement with the result of numerical simulations [Fig. \ref{Chern}(d)]. At $\Lambda=0$, the manifold of the spherical parameter space contains degeneracy at the center [Fig. \ref{Chern}(b)], indicating that the simulated Hamiltonian is in the
Maxwell metal phase and the extracted Chern numbers $|C_{\pm}|\approx2$. Moving
degeneracy along $R_{z}$ axis by varying $\Lambda$ [as illustrated in Fig.
\ref{Chern}(e)] is equivalent to deforming the manifold in the language of
topology. When $|\Lambda|<1$, $|C_{\pm}|\approx2$ indicates that the degeneracy
still lies inside spherical manifold. When the degeneracy is moving outside
the parameter sphere for $|\Lambda|>1$, $|C_{\pm}|\approx0$ indicates that the
system becomes a trivial insulator. Hence, topological phase transitions
occur at $|\Lambda|=1$, where $C_{\pm}$ will jump between discrete
values. Our measurements capture essential features of the
theoretical prediction [Fig. \ref{Chern}(e)]. It is noticed that the
transition of $C_{\pm}$ is not abrupt at the critical points, which is
mainly due to the finite decoherence time of the transmon. The simulation
results (solid line) by considering the decoherence time of the transmon
agree well with the experimental data \cite{Supp}.

In summary, we have explored essential physics of the momentum space Hamiltonian corresponding to topological Maxwell metal bands with a superconducting qutrit, which can be generalized to other artificial systems, including photonic crystals \cite{Lu,Chan} and trapped ions \cite{Ion}. A next study in this Maxwell system is to simulate complex relativistic quantum dynamics of spin-1 particles beyond the Dirac dynamics \cite{Castro}, such as super-Klein tunneling \cite{Fang} and double-Zitterbewegung oscillations. By using more energy levels in the superconducting artificial atom, one can emulate topological bands with higher-spin relativistic dispersions, such as spin-3/2 Rarita-Schwinger-Weyl semimetals \cite{Liang}. Furthermore, by coupling individual superconducting qutrits properly, one can extend the system to explore the topological phase transition induced by the qutrit-qutrit interaction, similar to that observed in the qubit-qubit interacting system \cite{Roushan}, even in principle, to implement the celebrated topological Haldane phase of interacting spin-1 quantum chain \cite{Haldane}.

\acknowledgments
This work was supported by the NKRDP of China (Grant No.
2016YFA0301800), the NSFC (Grants No. 11604103, No. 11474153, and No.
91636218), the NSF of Guangdong Province (Grant No. 2016A030313436), and the
Startup Foundation of SCNU.

X. T. and D.-W. Z contributed equally to this work.



\end{document}